# Time-Varying Coronary Artery Deformation: A Dynamic Skinning Framework for Surgical Training

Shuo Wang
Department of Engineering Physics, Key Laboratory of Particle and Radiation Imaging, Ministry of Education, Tsinghua University, Beijing, China
ORCID: https://orcid.org/0009-0008-6187-0401

Tong Ren
Department of Adult Cardiac Surgery, Senior Department of Cardiology, The Six medical center of PLA General Hospital, Beijing, China
Chinese PLA Medical School, Beijing, China

Nan Cheng
Department of Adult Cardiac Surgery, Senior Department of Cardiology, The Six medical center of PLA General Hospital, Beijing, China

Rong Wang*
Department of Adult Cardiac Surgery, Senior Department of Cardiology, The Six medical center of PLA General Hospital, Beijing, China
Corresponding Author
Email: wangrongd@126.com
Tel: +86-15801371359

Li Zhang*
Department of Engineering Physics, Key Laboratory of Particle and Radiation Imaging, Ministry of Education, Tsinghua University, Beijing, China
Corresponding Author
Email: zli@mail.tsinghua.edu.cn
Tel: +86-13910686306

# Time-Varying Coronary Artery Deformation: A Dynamic Skinning Framework for Surgical Training

**Abstract**

**Purpose** This study proposes a novel anatomically-driven dynamic modeling framework for coronary arteries using skeletal skinning weights computation, aiming to achieve precise control over vessel deformation while maintaining real-time performance for surgical simulation applications.

**Methods** We developed a computational framework based on biharmonic energy minimization for skinning weight calculation, incorporating volumetric discretization through tetrahedral mesh generation. The method implements temporal sampling and interpolation for continuous vessel deformation throughout the cardiac cycle, with mechanical constraints and volume conservation enforcement. The framework was validated using clinical datasets from 5 patients, comparing interpolated deformation results against ground truth data obtained from frame-by-frame segmentation across cardiac phases.

**Results** The proposed framework effectively handled interactive vessel manipulation. Geometric accuracy evaluation showed mean Hausdorff distance of 4.96 ± 1.78 mm and mean surface distance of 1.78 ± 0.75 mm between interpolated meshes and ground truth models. The Branch Completeness Ratio achieved 1.82 ± 0.46, while Branch Continuity Score maintained 0.84 ± 0.06 (scale 0-1) across all datasets. The system demonstrated capability in supporting real-time guidewire-vessel collision detection and contrast medium flow simulation throughout the complete coronary tree structure.

**Conclusion** Our skinning weight-based methodology enhances model interactivity and applicability while maintaining geometric accuracy. The framework provides a more flexible technical foundation for virtual surgical training systems, demonstrating promising potential for both clinical practice and medical education applications. The code is available at https://github.com/ipoirot/DynamicArtery.

# Introduction

Cardiovascular diseases, particularly coronary artery disease, remain one of the leading causes of mortality worldwide [1]. Dynamic modeling of coronary arteries has become increasingly important across multiple domains, including computational fluid dynamics (CFD), medical visualization, and virtual surgical simulation [2,3]. This multi-disciplinary significance stems from the complex nature of coronary arteries and their critical role in cardiac function.

The application of CFD simulations in coronary artery analysis has demonstrated remarkable capabilities in providing temporal and spatial distributions of pressure and flow patterns. These simulations enable detailed analysis of wall shear stress oscillations and help understand atherosclerotic plaque deposition mechanisms [4,5]. Recent studies have shown that abnormal hemodynamics, particularly low and oscillatory wall shear stress, plays a crucial role in the initiation and progression of atherosclerosis [6]. Beyond computational analysis, dynamic coronary modeling has revolutionized medical visualization by enabling 4D visualization of coronary motion during cardiac cycles [7,8].

In the realm of virtual surgical simulation, dynamic modeling has significantly improved surgical training and pre-operative planning [9]. Studies have demonstrated that virtual reality-based surgical simulation can effectively enhance surgical skills and reduce procedural complications [10,11]. The ability to simulate realistic vessel deformation has become particularly important for optimizing stent placement procedures and predicting post-intervention outcomes [12].

However, the implementation of accurate dynamic coronary modeling faces substantial challenges due to the complex anatomical characteristics of coronary arteries. The cyclic nature of cardiac motion, combined with multiple branches and bifurcations, creates intricate patterns of vessel movement and deformation. Furthermore, the variable vessel diameter and curvature, along with complex interactions with surrounding tissues, demand sophisticated modeling approaches that can capture these nuanced behaviors while maintaining computational efficiency.

Technical challenges in dynamic coronary modeling include balancing high-quality mesh generation with real-time performance requirements, while maintaining vessel elasticity representation and compatibility with clinical workflows. Current approaches typically focus on isolated aspects like static reconstruction or simplified dynamics, lacking comprehensive solutions for multiple applications. Although MeshMorphing techniques show promise [13], a more integrated approach is needed for applications ranging from CFD to surgical simulation.

Modern medical imaging technologies provide rich data sources for dynamic modeling through CT angiography, magnetic resonance imaging, and intravascular ultrasound. However, transforming this wealth of data into useful dynamic models presents significant challenges in maintaining visual quality during animation and ensuring smooth motion transitions. The handling of incomplete or noisy data remains a persistent challenge, particularly in cases where vessel tracking may be temporarily lost during certain cardiac phases.

To address the challenges in transforming medical imaging data into useful dynamic models, we propose a novel framework that combines advanced image processing with skeletal skinning approaches. Our key innovations include:

(1) A robust dynamic tracking method that maintains natural vessel deformation through intelligent interpolation, even with partial data loss

(2) Generation of topologically consistent CFD meshes across all cardiac phases, significantly reducing interpolation errors and computational time

(3) Real-time interaction capability supporting surgical simulation while maintaining consistent

topology throughout the cardiac cycle

This integrated approach bridges the gap between theoretical modeling and practical clinical applications, providing a robust foundation for medical education, surgical training, and clinical decision support. The framework effectively handles data from various imaging modalities while ensuring high-quality visualization and smooth motion transitions.

## Methods

### *Construction of Coronary Arterial Tree Model*

The construction of an anatomically accurate coronary arterial tree model involves multiple sophisticated steps, beginning with medical image acquisition and proceeding through several processing stages. The process integrates both imaging data and anatomical knowledge to ensure physiological accuracy.

### *Medical Image Acquisition and Preprocessing*

Cardiac CT imaging was performed using a Revolution Apex scanner (GE Healthcare, Milwaukee, WI, USA) during end-expiratory breath-hold with the patient in normal sinus rhythm. Image reconstruction was conducted across 20 temporal phases of the cardiac cycle, spanning from -5% to 106% of the R-R interval. The acquisition parameters included an isotropic slice thickness and increment of 0.625 mm, with an in-plane spatial resolution of 0.23 × 0.23 mm. The mean effective radiation dose, calculated in accordance with European Guidelines for Quality Criteria for Computed Tomography, was 10.49 mSv.

### *Vessel Segmentation and Centerline Extraction*

The coronary artery segmentation is performed on end-diastolic CTA images (Fig.1a), where vessels are maximally dilated with optimal contrast enhancement and minimal motion artifacts. The

Submitted to International Journal of Computer Assisted Radiology and Surgery

process employs the Frangi vesselness filter [14], which is based on the analysis of the Hessian matrix eigenvalues to detect tubular structures. For a given scale $\sigma$, the vesselness measure $V_0(s)$ is defined as:

$$V_0(s) = \begin{cases} 0 & \text{if } \lambda_2 > 0 \text{ or } \lambda_3 > 0 \\ \left(1 - \exp\left(-\frac{R_A^2}{2\alpha^2}\right)\right)\exp\left(-\frac{R_B^2}{2\beta^2}\right)\left(1 - \exp\left(-\frac{S^2}{2c^2}\right)\right) & \text{otherwise} \end{cases} \quad (1)$$

where $R_A = \left|\frac{\lambda_2}{\lambda_3}\right|$ is ratio for distinguishing between plate and line structures, $R_B = \frac{|\lambda_1|}{\sqrt{\lambda_2 \lambda_3}}$ is ratio for distinguishing blob-like structures; $\lambda_1$, $\lambda_2$, and $\lambda_3$ are eigenvalues of the Hessian matrix; $S = \sqrt{\lambda_1^2 + \lambda_2^2 + \lambda_3^2}$ is Frobenius norm of the Hessian, $\alpha$, $\beta$, and $c$ are thresholds controlling the sensitivity of the filter. The final vesselness measure $V$ is obtained by computing the maximum response over a range of scales:

$$V = max_{\sigma_{min} < \sigma < \sigma_{max}} V_0(s) \quad (2)$$

where $\sigma$ represents the scale parameter (in mm), $\sigma_{min}$ and $\sigma_{max}$ define the range of vessel sizes to be detected (typically $\sigma_{min} = 0.5$ mm and $\sigma_{max} = 2.5$ mm for coronary arteries).

Following segmentation, centerline extraction is performed using a fast marching algorithm combined with backtracking (Fig.1c). The centerline coordinates are represented as a series of points $P_i(x_i, y_i, z_i)$, where each point carries additional information about vessel radius $r$, and bifurcation relationships [7].

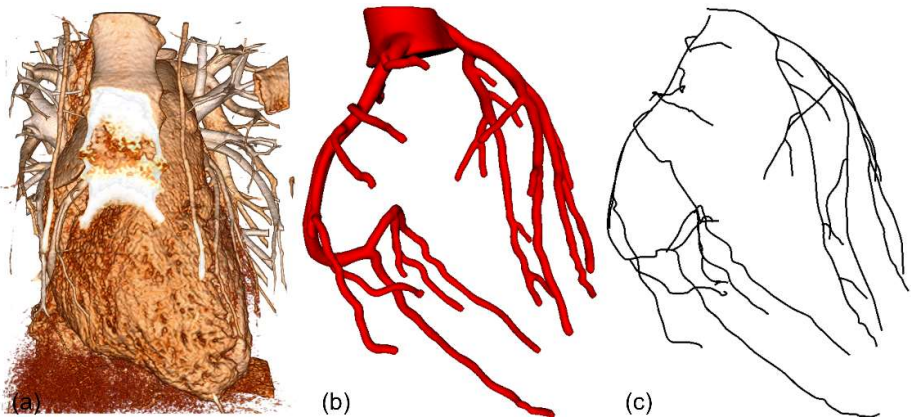
(a) (b) (c)

Submitted to International Journal of Computer Assisted Radiology and Surgery

**Fig. 1** Coronary artery segmentation workflow. (a) Volume rendering of cardiac CTA data. (b) Segmented coronary artery mask using Frangi vesselness filter. (c) Extracted vessel centerlines using fast marching algorithm

## *Topology Analysis and Branch Identification*

The arterial tree topology is analyzed using a hierarchical approach. Starting from the main coronary arteries (left main, left anterior descending, left circumflex, and right coronary artery), branches are identified and classified based on their anatomical positions and geometric features. As shown in Fig. 2, branch points are characterized by their bifurcation angles and cross-sectional area ratios, following Murray's law for physiological branching patterns [15]:

$$r_p^3 = r_{d1}^3 + r_{d2}^3 \qquad (3)$$

where $r_p$ is the parent vessel radius, $r_{d1}$ and $r_{d2}$ are daughter vessel radii.

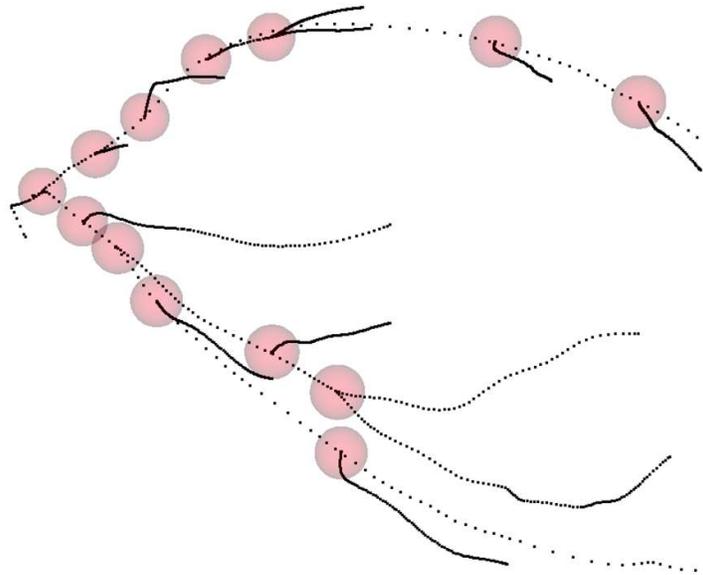

**Fig. 2** Branch point detection in left coronary artery. Red spheres indicate the detected bifurcation points along the vessel centerlines

## *Surface Mesh Generation*

The vessel surface is reconstructed using a hybrid approach combining implicit surface fitting and explicit mesh generation. Cross-sections are positioned along the parameterized centerline using the Frenet-Serret trihedron (Fig. 3a). For each cross-section, the patch pattern is generated using a central point and an arbitrary shaped outer contour in the trihedron normal-binormal plane, following the approach of Vukicevic et al.[16]. These cross-sectional patches are then connected using B-spline interpolation to create a smooth surface representation (Fig. 3b). The surface mesh is generated using an advancing front method, with mesh density adapted according to local geometric features:

$$h(x) = h_{min} + (h_{max} - h_{min})e^{-\alpha\kappa(x)} \qquad (4)$$

where $h(x)$ is the local mesh size, $\kappa(x)$ is the local curvature, and $\alpha$ is a control parameter.

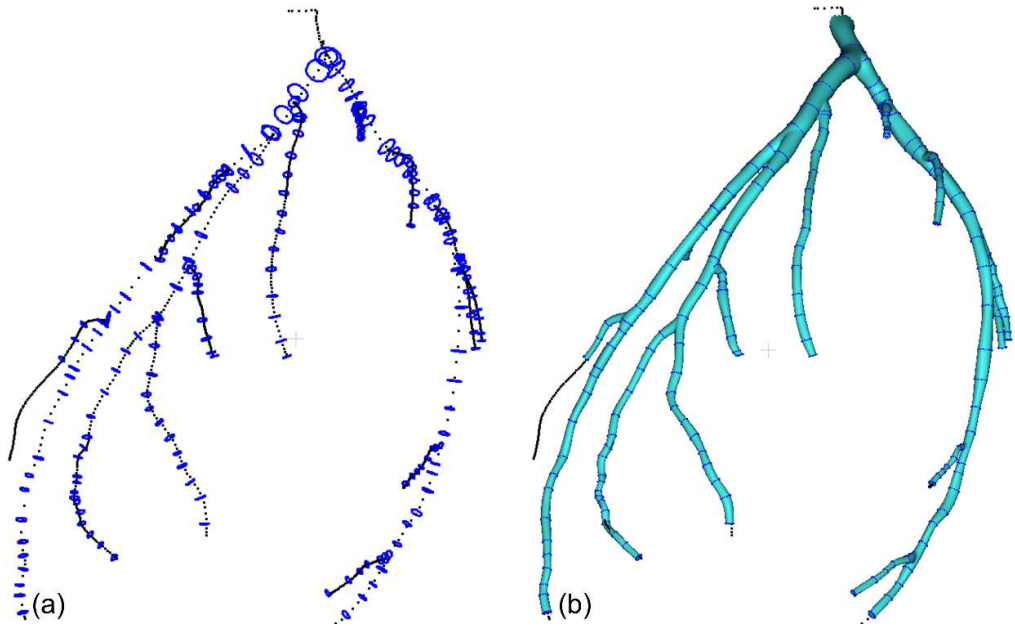

**Fig. 3** Surface reconstruction of coronary arteries. (a) Cross-sections with arbitrary shaped contours extracted along the centerline using Frenet-Serret trihedron. (b) Final smooth surface representation generated by B-spline interpolation between cross-sectional patches

### *Coronary Artery Skinning Weight Calculation*

The computation of skinning weights for coronary artery deformation follows a comprehensive

Submitted to International Journal of Computer Assisted Radiology and Surgeryanatomical framework that ensures both robustness and physiological fidelity. The process begins with a preprocessing stage that handles the input coronary artery mesh $M = (V, F)$, where a cleaning operation is performed to ensure topological consistency and remove artifacts from medical imaging reconstruction. This operation produces a cleaned vessel mesh $M_{clean}$ that serves as the foundation for subsequent computations.

The cornerstone of our approach lies in the volumetric discretization of the arterial wall through tetrahedral mesh generation. We employ constrained Delaunay tetrahedralization to create a volumetric representation $T = (V_T, T_T)$ that encompasses both the vessel wall geometry and the centerline structure. This process carefully preserves the anatomical characteristics of coronary arteries while ensuring that each tetrahedron maintains a minimum volume threshold of $10^{-7}$ to prevent numerical instabilities in subsequent computations.

The weight computation follows a biharmonic energy minimization framework, where we solve for weights $W$ that minimize the Laplacian energy while satisfying vessel wall constraints. The optimization problem is formulated as:

$$\min_{w_b} \int_\Omega \|\Delta w_b(v)\|^2 dv \tag{5}$$

subject to boundary conditions that enforce unity weights at centerline locations and zero weights at vessel wall boundaries, where $w_b$ represents the skinning weight function for the b-th centerline segment, $\Delta$ is the Laplacian operator measuring second spatial derivatives, $v$ denotes a point in 3D space within the vessel wall domain $\Omega$, and $dv$ is the differential volume element. This continuous optimization is discretized and solved using an active set method with careful constraint handling to maintain both boundedness ($0 \leq w_{b(v)} \leq 1$) and partition of unity ($\sum_b w_{b(v)} = 1$).

The numerical solution employs an efficient sparse matrix formulation, where the discrete bi-

Laplacian operator $L$ is constructed to capture the geometric properties of the arterial wall mesh. The boundary conditions are encoded through constraint matrices, leading to the discrete optimization problem:

$$\min_W tr(W^T L W) \quad subject\ to: BW = BC \tag{6}$$

where $W$ is the weight matrix (columns for centerline segments, rows for mesh vertices), $L$ is the discrete bi-Laplacian operator matrix, $tr(\cdot)$ denotes the matrix trace operator, and the constraint $BW = BC$ encodes boundary conditions, with $B$ being the constraint matrix and $BC$ containing the boundary values.

To ensure numerical stability and physiological validity, the computed weights undergo a normalization process that enforces the partition of unity constraint while preserving the relative influence of each centerline segment. This is achieved through row-wise normalization:

$$w_{i,j} = \frac{w_{i,j}}{\sum_k w_{i,k}} \tag{7}$$

where $i$ for vertex index on the vessel wall mesh, $j$ for centerline segment index, and $k$ as the summation variable iterating through all centerline segments.

The weight distribution method ensures smooth variation across vessel walls while maintaining local control and non-negative weights, crucial for preserving vessel wall continuity during cardiac motion-induced deformation. This mathematically rigorous and physiologically accurate approach effectively handles complex arterial structures through efficient sparse matrix operations, outperforming traditional weight computation methods in capturing the intricate relationships between centerline segments and vessel wall geometry. The coronary artery deformation process is demonstrated in Figure 4. Starting with the original mesh and deformation skeletal nodes (Figure 4a), the artery underwent two successive deformations (Figure 4b,c) using our skinning weight calculation method. The complete deformation validation process can be visualized as shown in the animation (Online Resource 3).

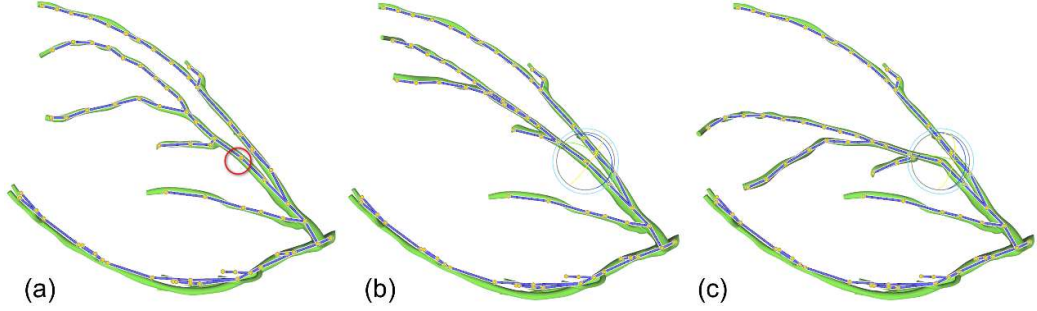

**Fig. 4** Coronary artery deformation using skinning weight calculation. (a) Original left coronary artery mesh with deformation skeletal nodes marked in red circles. (b) Result after first deformation. (c) Result after second deformation

## *Dynamic Sequence Generation for Coronary Artery Motion*

Dynamic sequence generation for coronary artery motion is a comprehensive process that creates realistic vessel deformation patterns throughout the cardiac cycle. The process begins with temporal sampling and interpolation, where the cardiac cycle is divided into 20 temporal phases, spanning from -5% to 106% of the R-R interval. Key poses at these phases serve as anchors, while intermediate frames between captured phases are generated through temporal interpolation using the formula:

$$P(t) = (1-t)P_1 + tP_2 \tag{8}$$

where $P_1$ and $P_2$ represent consecutive key poses, and t varies between 0 and 1.

The centerline evolution forms the backbone of the motion sequence, described by a time-dependent curve

$$C(s,t) = (x(s,t), y(s,t), z(s,t)) \tag{9}$$

where $s$ represents the arc-length parameter and $t$ denotes time. This motion must satisfy continuity conditions, ensuring that $\frac{\partial C}{\partial s}$ remains continuous for smoothness, while $\frac{\partial^2 C}{\partial s^2}$ captures the changes in curvature. The vessel cross-section undergoes simultaneous deformation, modeled as an ellipse with time-varying parameters following the equation:

$$\frac{x^2}{a(t)^2} + \frac{y^2}{b(t)^2} = 1 \tag{10}$$

where $a(t)$ and $b(t)$ represent the major and minor axis lengths. The cross-sectional area follows the volume conservation law:

$$\pi a(t)b(t) = A_0\big(1 + \epsilon_r(t)\big) \tag{11}$$

where $A_0$ is the reference cross-sectional area and $\epsilon_r(t)$ represents the small radial strain that accounts for the vessel's slight compressibility during the cardiac cycle, typically within ±5% range.

The motion sequence must adhere to various anatomical and mechanical constraints. These include maximum bending angles $\theta_{max}$, minimum radius of curvature $R_{min}$, strain limits $\epsilon_{max}$, and the fundamental stress-strain relationship:

$$\sigma = E\epsilon \tag{12}$$

where $\sigma$ is the stress, $E$ is Young's modulus, and $\varepsilon$ is the strain.

The velocity field, computed as $v(x,t) = \frac{\partial P}{\partial t}$, undergoes regularization through the equation:

$$\nabla^2 v + \lambda \frac{\partial v}{\partial t} = 0 \tag{13}$$

where $\lambda$ serves as a smoothing parameter to ensure fluid motion transitions.

To maintain realistic motion sequences, we enforce temporal coherence through position constraints between consecutive frames: $\|P(t + \Delta t) - P(t)\| \leq \delta$, where $\delta$ is the maximum allowed displacement threshold between adjacent frames to ensure smooth motion transitions. Additionally, periodic boundary conditions ensure that $P(T) = P(0)$, where $T$ represents the cardiac cycle period.

The quality of the generated sequence is evaluated through temporal consistency $E_{temp} = \sum_{t=1}^{N}\|P_t - P_{t-1}\|^2$ and spatial smoothness:

$$E_{spatial}(t) = \sum_{i=1}^{M}\|\Delta P_{t,i}\|^2 \tag{14}$$

where $\Delta P_{t,i} = \frac{1}{|N(i)|}\sum_{j\in N(i)}(P_{t,j} - P_{t,i})$ represents the discrete Laplacian operator with $N(i)$ denoting the neighborhood of point $i$. The total spatial energy is then computed as $E_{spatial} = \sum_{t=1}^{N} E_{spatial}(t)$. These terms are combined in combined in a total energy minimization framework $E_{total} = \omega_1 E_{temp} + \omega_2 E_{spatial}$.

This comprehensive approach results in a continuous, physically plausible representation of coronary artery motion throughout the cardiac cycle, as shown in Figure 5. The generated sequences effectively balance mathematical precision with physiological realism while maintaining computational efficiency. The resulting dynamic sequences provide valuable insights into vessel behavior and serve as essential tools for both visualization and biomechanical analysis in clinical applications. The accuracy of these sequences directly impacts the quality of subsequent analyses and simulations, making the generation process a critical component in understanding coronary artery dynamics.

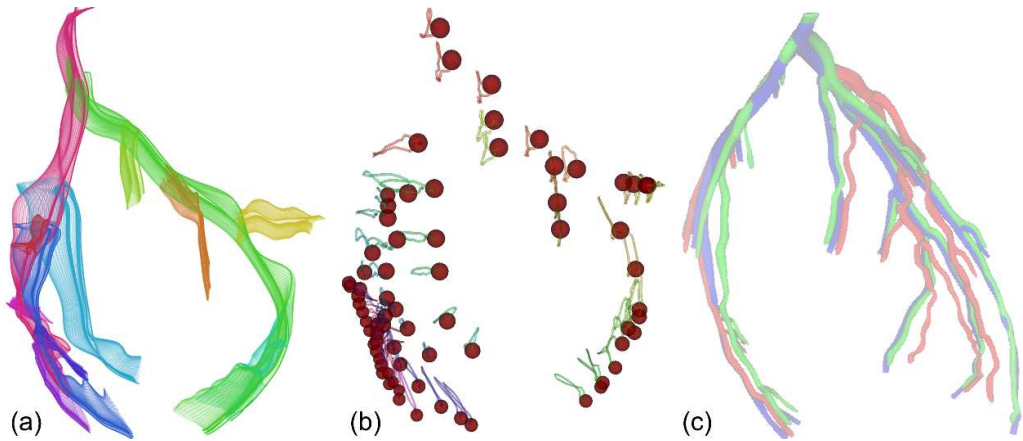

**Fig. 5** Visualization of coronary artery dynamic sequence generation. (a) Temporal evolution of left coronary artery centerlines, where different colors denote distinct arterial branches with preserved geometric continuity. (b) Trajectory analysis depicting initial anatomical positions (red spheres) and their corresponding motion paths (semi-transparent tubes) under prescribed displacement constraints. (c) Multi-phase visualization of left coronary artery meshes at three temporal positions within the cardiac cycle, demonstrating the dynamic deformation of vessel cross-sections

# Results

## *Validation of Dynamic Mesh Generation*

To rigorously validate the accuracy of our proposed dynamic mesh generation framework, we conducted comprehensive experiments on five independent 4D CT datasets. As shown in Figure 6, we present a detailed analysis of one representative case comprising 20 temporal phases, while similar validation procedures were performed across all five datasets. The validation strategy employed a systematic approach where meshes at even-numbered phases (0, 2, 4, ...) were reconstructed using direct tracking-based skinning weights, while odd-numbered phases (1, 3, 5, ...) were generated through our interpolation method utilizing adjacent frame information. Notably, while traditional segmentation-based methods often suffer from incomplete branch reconstruction due to motion artifacts across different cardiac phases, our skinning weights-based approach demonstrates significant advantages in maintaining coronary tree completeness. By using end-diastolic phase as the reference, where coronary arteries are most clearly visible and complete, our method successfully preserves the integrity of the branching structure throughout the cardiac cycle. This approach effectively addresses a common limitation in conventional segmentation methods, where motion artifacts frequently result in discontinuous or missing branches in various phases.

For quantitative assessment, we established a ground truth dataset by performing frame-by-frame segmentation and reconstruction across all 20 phases, represented as green meshes $M^*(t)$. The interpolated meshes at odd-numbered phases, denoted as $M_I(t)$, were rendered in red for visual comparison. This experimental design enables direct evaluation of our interpolation method's accuracy at intermediate phases where direct tracking data is unavailable.

The geometric accuracy of the interpolated meshes was evaluated using the Hausdorff distance (HD) metric:

$$d_H(M_I, M^*) = \max\left(\sup_{x \in M_I} \inf_{y \in M^*} d(x,y), \sup_{y \in M^*} \inf_{x \in M_I} d(x,y)\right) \tag{15}$$

where $d_H$ represents the Hausdorff distance, $M_I$ denotes the interpolated mesh, $M^*$ represents the ground truth mesh, $d(x,y)$ is the Euclidean distance between points $x$ and $y$, $sup$ denotes the supremum (least upper bound), and $inf$ denotes the infimum (greatest lower bound). Additionally, we computed the mean surface distance (MSD) to assess overall mesh similarity:

$$MSD(M_I, M^*) = \frac{1}{|M_I|} \sum_{x \in M_I} \min_{y \in M^*} d(x,y) \tag{16}$$

where $|M_I|$ represents the total number of vertices in the interpolated mesh, and min denotes the minimum distance from a point $x$ on $M_I$ to any point y on $M^*$.

To specifically evaluate the topological completeness of coronary trees, we introduced two complementary metrics: Branch Completeness Ratio (BCR) and Branch Continuity Score (BCS). The BCR quantifies the preservation of branch numbers while BCS assesses the structural continuity:

$$BCR = \frac{N_I}{N_R}, \quad BCS = \frac{1}{N_I} \sum_{i=1}^{N_I} \left(\frac{L_i}{L_R^i}\right) \tag{17}$$

where $N_I$ and $N_R$ represent the number of detected branches in the interpolated and reference meshes respectively, $L_i$ denotes the length of the i-th branch in the interpolated mesh, and $L_R^i$ is its corresponding length in the reference mesh. Together, these metrics provide a comprehensive assessment of both topological preservation and structural continuity of the coronary tree during the interpolation process.

The quantitative evaluation results are summarized in Table 1. Part A presents a detailed phase-by-phase analysis for the representative case (Dataset #1) corresponding to the visual results shown in Figure 6. This comprehensive temporal analysis demonstrates the consistency of our interpolation method across different cardiac phases. Part B summarizes the performance across all five datasets, where each dataset value represents the mean across all interpolated phases. The evaluation metrics include geometric accuracy measures (HD and MSD) and topological preservation measures (BCR and BCS). These results quantitatively validate that our proposed method achieves both accurate geometric interpolation and robust topological preservation of the coronary tree structure throughout the cardiac cycle.

Based on the quantitative evaluation results shown in Table 1 Part A, our interpolation method demonstrates robust performance across different cardiac phases. The HD exhibits a mean of $5.42 \pm 2.60$ mm, while the MSD shows better consistency with $1.99 \pm 1.03$ mm, indicating generally reliable geometric accuracy of the interpolated meshes. The BCR ($1.95 \pm 1.18$) consistently exceeds 1.0, suggesting that our method effectively preserves and sometimes enhances branch detection compared to the reference meshes, addressing the common challenge of incomplete branch reconstruction in traditional segmentation-based approaches. The BCS maintains a high average of $0.82 \pm 0.07$, with values ranging from 0.71 to 0.95, demonstrating stable structural continuity preservation throughout the cardiac cycle. Notably, phases with lower geometric distances (e.g., Phase 15 with HD = 0.79 mm, MSD = 0.11 mm) achieve higher continuity scores (BCS = 0.95), confirming the correlation between geometric accuracy and topological preservation.

The complete processing pipeline for a 4D CT dataset is demonstrated in the animation (Online Resource 1), showing the key steps of segmentation, temporal interpolation, and visualization of the dynamic coronary tree model.

**Table 1** Quantitative evaluation metrics for coronary artery mesh interpolation across five 4D CT datasets.

**Part A**: Individual phase analysis for Dataset #1 (Representative case)

| Phase | HD (mm) | MSD(mm) | BCR | BCS |
|---|---|---|---|---|
| 01 | 5.75 | 1.49 | 2.5 | 0.83 |
| 03 | 7.70 | 3.39 | 1.66 | 0.76 |
| 05 | 6.79 | 2.75 | 5.0 | 0.71 |
| 07 | 8.20 | 2.97 | 2.5 | 0.75 |
| 09 | 2.27 | 0.93 | 1.25 | 0.89 |
| 11 | 5.88 | 2.36 | 1.0 | 0.85 |
| 13 | 5.00 | 2.55 | 1.66 | 0.82 |
| 15 | 0.79 | 0.11 | 1.0 | 0.95 |
| 17 | 3.47 | 0.97 | 1.66 | 0.87 |
| 19 | 8.34 | 2.39 | 1.25 | 0.77 |
| Mean | 5.42 | 1.99 | 1.95 | 0.82 |
| STD | 2.60 | 1.03 | 1.18 | 0.07 |

**Part B**: Summary statistics across all datasets

| Dataset | HD (mm) | MSD (mm) | BCR | BCS |
|---|---|---|---|---|
| #1 | 6.23 | 2.16 | 2.00 | 0.81 |
| #2 | 3.89 | 1.28 | 1.50 | 0.88 |
| #3 | 7.16 | 2.89 | 2.50 | 0.75 |
| #4 | 4.57 | 1.56 | 1.80 | 0.84 |
| #5 | 2.95 | 0.99 | 1.30 | 0.90 |
| Mean | 4.96 | 1.78 | 1.82 | 0.84 |
| STD | 1.78 | 0.75 | 0.46 | 0.06 |

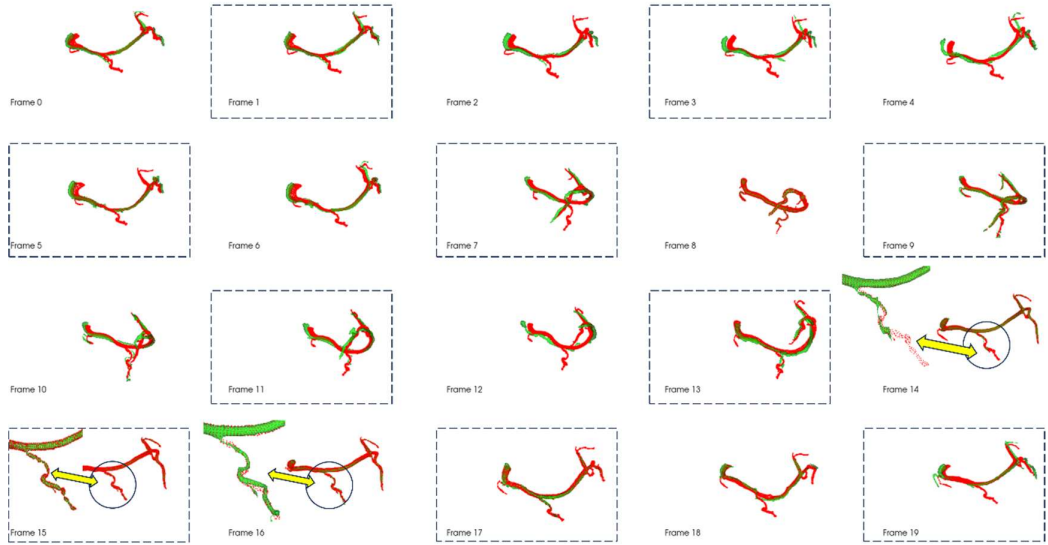

**Fig. 6** Visual comparison of coronary artery meshes over a cardiac cycle. Twenty phases (0-19) arranged in 4×5 grid show interpolated meshes (red) overlaid with ground truth segmentations (green). Dashed-border frames (1,3,5,7,...) represent interpolated phases, while others (0,2,4,6,...) were reconstructed using direct tracking-based weights. Consistent mesh overlap demonstrates anatomical accuracy throughout the cycle. Magnified views of frames 14-16 show successful mesh reconstruction in frame 15 despite tracking loss in frame 14

## *Validation Through Application: PCI Simulation*

To demonstrate the practical utility of our dynamic mesh reconstruction framework, we integrated it into a coronary intervention simulation system (Fig. 7). This integration enables realistic catheterization training scenarios with accurate vessel deformation and fluid dynamics simulation throughout the cardiac cycle. The simulation framework incorporates two critical components: (1) real-time collision detection between the guide wire and dynamically deforming coronary arteries, as shown in Fig. 7, and (2) contrast medium flow simulation within the complete coronary tree structure (Fig. 8).

Our dynamic mesh generation method provides essential support for both aspects: For guide wire navigation simulation, we implemented a dense collision detection algorithm between the discrete guide wire elements and the temporally varying coronary mesh. The complete topological structure maintained by our method ensures continuous collision response as the guide wire navigates through different branches. The collision detection is performed using a hierarchical bounding volume structure that is updated according to the mesh deformation at each cardiac phase, enabling efficient real-time performance while maintaining accuracy. The contrast medium simulation leverages our preserved vessel topology to accurately model fluid dynamics within the coronary tree. We adopted a particle-based fluid simulation approach where the complete and continuous vessel structure is essential for realistic contrast medium propagation. The dynamic mesh sequence provides accurate boundary conditions for the fluid simulation, allowing the contrast medium to flow naturally through the deforming vessels while respecting the cardiac motion.

The simulation results demonstrate that our dynamic mesh generation method effectively supports both the mechanical aspects (guide wire-vessel interaction) and fluid dynamics (contrast medium flow) required for realistic coronary intervention simulation. The preserved topological completeness and geometric accuracy of the vessel structure throughout the cardiac cycle prove crucial for achieving high-fidelity simulation outcomes. Real-time coronary intervention simulation as shown in the animation (Online Resource 2).

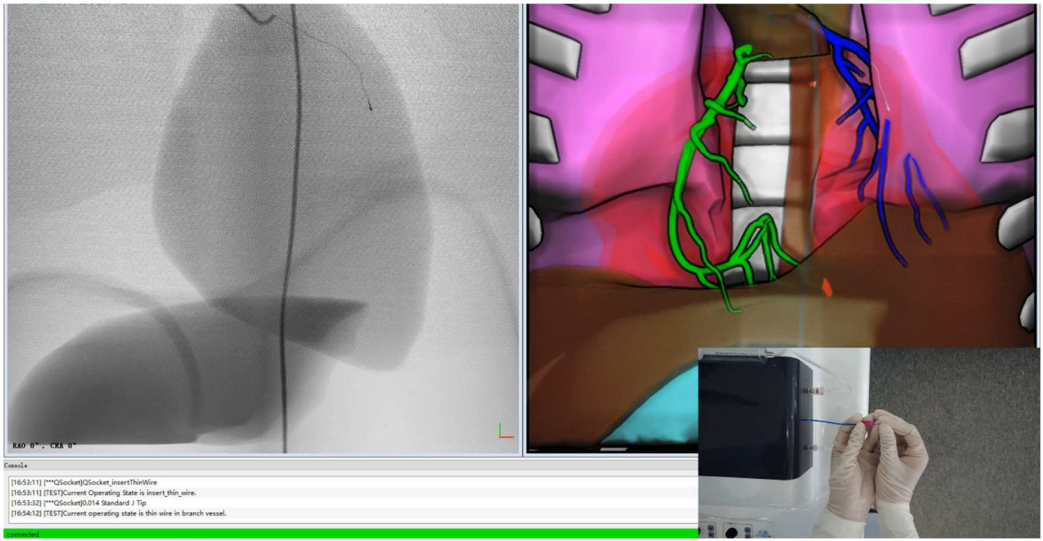

**Fig. 7** Real-time coronary intervention simulation. Left: Simulated X-ray fluoroscopic view showing the results of real-time collision detection and physical simulation of guidewire-vessel interaction. Right: Corresponding 4D anatomical scene with dynamic vessel deformation. Bottom right: PCI simulator hardware configuration

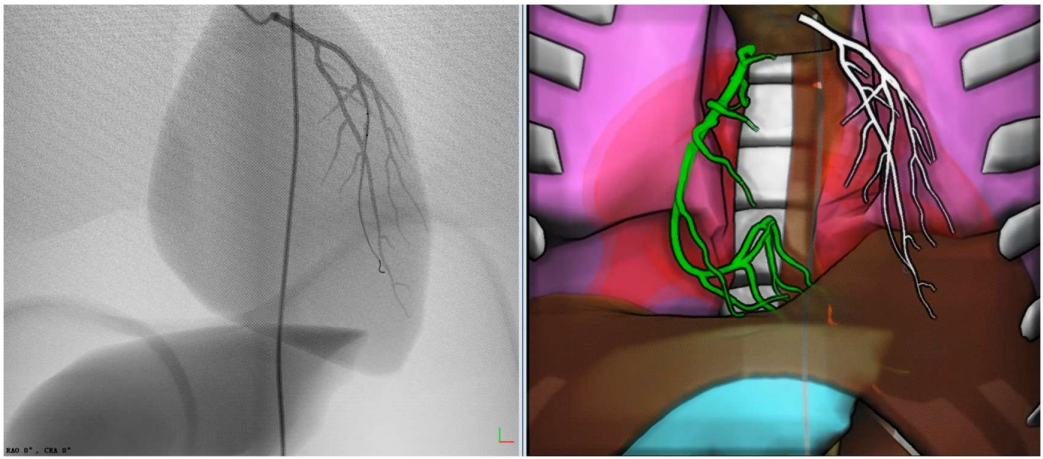

**Fig. 8** Simulated LCA contrast injection. The fluoroscopic sequence demonstrates dynamic contrast medium opacification throughout the left coronary tree during the cardiac cycle

## Discussion

Our skeletal skinning weights-based dynamic modeling framework for coronary arteries demonstrates significant advantages over recent image registration-based methods [17]. Our approach, utilizing biharmonic energy minimization for deformation weight computation, enables precise vessel morphology control while extending traditional CFD analysis capabilities.

The framework represents a significant advancement in virtual surgical training applications. The skinning weights-based approach enables clinicians to modify vessel morphology through control points, rapidly adapting to individual patient anatomy. Tetrahedral mesh generation ensures robust weight computation, while the biharmonic energy minimization framework guarantees smooth weight distribution across vessel walls.

Our dynamic mesh generation framework has been successfully integrated into real-time surgical simulation, supporting guidewire-vessel collision detection and contrast medium flow simulation, providing effective tools for pre-operative simulation and surgical planning.

While the current framework shows promising results, certain limitations and future directions warrant consideration. Computational efficiency requires optimization, particularly for real-time deformation of complex branching structures, and further validation is needed for deformation accuracy across various pathological conditions. While Psiuk-Maksymowicz et al.'s approach excels in CFD analysis, our method's capability for interactive manipulation and real-time deformation makes it particularly suitable for virtual reality-based surgical training and education. This complementary advancement suggests that future developments might benefit from combining both approaches to create more comprehensive solutions for cardiovascular modeling and simulation.

## Conclusions

This paper presents a novel dynamic modeling framework for coronary arteries based on anatomically-driven skeletal skinning weights. Through the implementation of biharmonic energy minimization and volumetric discretization, we have demonstrated the capability to generate physically plausible vessel deformations that maintain anatomical accuracy throughout the cardiac cycle. Unlike traditional approaches that focus solely on computational fluid dynamics, our framework enables real-time manipulation of vessel morphology while preserving mechanical constraints and volume conservation.

Experimental results have validated the framework's effectiveness in supporting both mechanical deformation and fluid dynamics simulation, particularly in handling complex vessel topologies while maintaining computational efficiency suitable for interactive applications. The integration with virtual reality-based surgical simulation platforms demonstrates the practical utility of our approach in medical education and pre-operative planning. While current limitations in computational optimization present opportunities for improvement, this work establishes a foundation for future developments in personalized cardiovascular modeling, contributing to the ongoing evolution of computational tools for cardiovascular medicine.

## Code availability

The code is available at the following link: https://github.com/ipoirot/DynamicArtery.

## Acknowledgements

This study was funded by National Natural Science Foundation of China (62031020, 62476286), National Key R&D Plan(2022YFB4700800), Beijing Science and Technology Plan Project (Z231100005923039).

## Author information

### Authors and Affiliations

**Department of Engineering Physics, Key Laboratory of Particle and Radiation Imaging, Ministry of Education, Tsinghua University, Beijing, China**
Shuo Wang & Li Zhang
**Department of Adult Cardiac Surgery, Senior Department of Cardiology, The Six medical center of PLA General Hospital. Fucheng Road, Haidian District,100048,Beijing, China**

Tong Ren, Nan Cheng & Rong Wang

## Corresponding authors

Correspondence to Li Zhang or Rong Wang.

## Ethics declarations

**Conflict of interest** The authors declare that they have no Conflict of interest.
**Ethical approval** The Ethics Committee of the General Hospital of the Chinese People's Liberation Army accepted this protocol (S2024-580). Informed consent was obtained from all individual participants included in the study.

## References

1. Virani SS, Alonso A, Benjamin EJ, Bittencourt MS, Callaway CW, Carson AP, Chamberlain AM (2021) Heart Disease and Stroke Statistics-2021 Update: A Report From the American Heart Association. Circulation 143:e254-e743

2. Taylor CA, Fonte TA, Min JK, Koo BK, Leipsic J, Pontone G (2013) Computational Fluid Dynamics Applied to Cardiac Computed Tomography for Noninvasive Quantification of Fractional Flow Reserve. J Am Coll Cardiol 61:2233-2241. https://doi.org/10.1016/j.jacc.2012.11.083

3. Morris PD, Ryan D, Morton AC, Lycett R, Lawford PV, Hose DR (2013) Virtual Fractional Flow Reserve From Coronary Angiography: Modeling the Significance of Coronary Lesions. JACC Cardiovasc Interv 6:149-157. https://doi.org/10.1016/j.jcin.2012.08.024

4. Chatzizisis YS, Coskun AU, Jonas M, Edelman ER, Feldman CL, Stone PH (2007) Role of Endothelial Shear Stress in the Natural History of Coronary Atherosclerosis and Vascular Remodeling. J Am Coll Cardiol 49:2379-2393. https://doi.org/10.1016/j.jacc.2007.02.059

5. Stone PH, Maehara A, Coskun AU, Maynard CC, Zaromytidou M, Siasos G (2018) Role of Low Endothelial Shear Stress and Plaque Characteristics in the Prediction of Nonculprit Major Adverse Cardiac Events. JACC Cardiovasc Imaging 11:462-471. https://doi.org/10.1016/j.jcmg.2017.01.031

6. Kwak BR, Bäck M, Bochaton-Piallat ML, Caligiuri G, Daemen MJ, Davies PF (2014) Biomechanical factors in atherosclerosis: mechanisms and clinical implications. Eur Heart J 35:3013-3020. https://doi.org/10.1093/eurheartj/ehu353

7. Antiga L, Piccinelli M, Botti L, Ene-Iordache B, Remuzzi A, Steinman DA (2008) An image-based modeling framework for patient-specific computational hemodynamics. Med Biol Eng Comput 46:1097-1112. https://doi.org/10.1007/s11517-008-0420-1

8. Updegrove A, Wilson NM, Merkow J, Lan H, Marsden AL, Shadden SC (2017) SimVascular: An Open Source Pipeline for Cardiovascular Simulation. Ann Biomed Eng 45:525-541. https://doi.org/10.1007/s10439-016-1762-8

9. Ujiie H, Kato T, Igami T, Fujiwara M, Sato M, Nagino M (2024) Developing a Virtual Reality Simulation System for Preoperative Planning of Robotic-Assisted Thoracic Surgery. J Clin Med 13:611-619. https://doi.org/10.3390/jcm13020611

10. Joseph FJ, Rai A, Chandra PS, Suri A, Kumar A, Kale SS (2023) Simulation training approaches in

intracranial aneurysm surgery-a systematic review. Neurosurg Rev 46:101-108. https://doi.org/10.1007/s10143-023-01995-5

11. Gamberini G, Tognarelli S, Saracino A, Menciassi A, Sinibaldi E, Ciuti G (2024) Design and preliminary validation of a high-fidelity vascular simulator for robot-assisted manipulation. Sci Rep 14:4779-4790. https://doi.org/10.1038/s41598-024-55351-8

12. Han Z, Li J, Singh M, Wu C, Liu C, Zhao X (2024) A review on organ deformation modeling approaches for reliable surgical navigation using augmented reality. Comput Assist Surg 29:e2357164. https://doi.org/10.1080/24699322.2024.2357164

13. Pons-Moll G, Romero J, Mahmood N, Black MJ (2015) Dyna: A model of dynamic human shape in motion. ACM Trans Graph 34:1-14. https://doi.org/10.1145/2766993

14. Frangi AF, Niessen WJ, Vincken KL, Viergever MA (1998) Multiscale vessel enhancement filtering. Med Image Comput Comput Assist Interv 1496:130-137. https://doi.org/10.1007/BFb0056195

15. Murray CD (1926) The physiological principle of minimum work: I. The vascular system and the cost of blood volume. Proc Natl Acad Sci USA 12:207-214

16. Vukicevic AM, Stepanovic NM, Jovicic GR, Apostolovic SR, Filipovic ND (2014) Three-dimensional reconstruction and NURBS-based structured meshing of coronary arteries from CT images. Comput Methods Programs Biomed 117:642-654

17. Psiuk-Maksymowicz K, Polewczyk A, Orlef A, Walczak E, Czerwiński M, Gąsior Z (2024) Methodology of generation of CFD meshes and 4D shape reconstruction of coronary arteries from patient-specific dynamic CT. Sci Rep 14:2201-2215. https://doi.org/10.1038/s41598-024-52398-5